\documentclass[a4paper]{jpconf}
\usepackage{graphicx}
\usepackage{amssymb}
\usepackage{wrapfig}
\usepackage{setspace}
\usepackage{siunitx}
\usepackage{amsmath}
\usepackage{amssymb}

\begin{document}
\title{Signal Readout Electronics for \textsc{LEGEND}-200}

\author{Michael Willers {\normalfont for the LEGEND collaboration}}

\address{Lawrence Berkeley National Laboratory, 1 Cyclotron Road, Berkeley, CA 94720 USA}

\ead{mwillers@lbl.gov}

\begin{abstract}
\textsc{LEGEND} (Large Enriched Germanium Experiment for Neutrinoless Double-Beta Decay) is a
ton-scale experimental program to search for neutrinoless double beta ($0 \nu \beta \beta$) decay using high-purity germanium detectors enriched in the isotope $^{76}$Ge. An observation of $0 \nu \beta \beta$ decay would unequivocally prove the existence of lepton number violation and provide insight into the nature of neutrino masses.

On-site construction for the first 200-kg phase (LEGEND-200) of the experiment will start at the Gran
Sasso underground laboratory (Laboratory Nazionali del Gran Sasso, Italy) in mid-2020. In order to achieve a sensitivity
of $10^{27}$ years on the $0 \nu \beta \beta$ half-life, ultra-clean low-noise signal readout electronics are essential as they impact both the background budget of the experiment and the efficiency of background rejection.
In this contribution, the current status of the signal readout electronics for LEGEND-200 is presented.
\end{abstract}

\section{Introduction}
Double beta ($2 \nu \beta \beta$) decay $(A,Z) \to (A,Z+2) + 2e^- + 2\overline{\nu}_e$ is a second order nuclear decay that has been observed in several isotopes such as \textsuperscript{76}Ge, \textsuperscript{130}Te, and \textsuperscript{136}Xe with half-lifes in the order of $\sim 10^{19} - 10^{21} \,\textrm{yr}$.
Neutrinoless double beta ($0 \nu \beta \beta$) decay $(A,Z) \to (A,Z+2) + 2e^-$ is forbidden in the Standard Model (SM) of particle physics as it violates lepton number conservation and has so-far not been observed experimentally. 
The observation of $0 \nu \beta \beta$ decay would therefore unequivocally prove the existence of lepton number violation, new physics beyond the SM, and provide insight into the nature of neutrino masses  (see e.g. \cite{DBD}).\\

\noindent Using high purity germanium (HPGe) detectors enriched in the isotope \textsuperscript{76}Ge to search for $0 \nu \beta \beta$ decay is one of the most promising experimental techniques: HPGe detectors are a commercially mature technology;they can be enriched to up to $\gtrsim 88 \%$ in \textsuperscript{76}Ge; they have excellent energy resolution and a high detection efficiency for $\beta \beta$ interactions. Very good background discrimination is  enabled by the spatial and timing information that can be extracted from HPGe detector signals. The primary signature for $0 \nu \beta \beta$ decay is a distinctive peak at the Q value ($Q_{\beta\beta} = 2039 \,\textrm{keV}$) of the decay.

\section{LEGEND}
The Large Enriched Germanium Experiment for Neutrinoless Double-Beta Decay (LEGEND) collaboration is developing a tonne-scale \textsuperscript{76}Ge experimental program with discovery potential at half-lifes $> 10^{28} \,\textrm{yr}$ \cite{LEGEND}. The first phase of the LEGEND program, LEGEND-200, is currently under construction and will use 200 kg of HPGe detectors enriched in the isotope \textsuperscript{76}Ge. Detector installation and commissioning at the experimental site at the Gran Sasso underground laboratory (Laboratori Nazionali del Gran Sasso, Italy) is scheduled to begin in mid-2020 with data taking starting in early 2021. LEGEND-200 is designed to achieve a background level of $0.6\,\textrm{cts} / \textrm{(FWHM t yr)}$ in the region of interest (ROI) around $Q_{\beta \beta}$ and is expected to reach a half-life sensitivity of $10^{27} \,\textrm{yr}$ after taking data for 5 years. Subsequent phases of the LEGEND program (LEGEND-1000) will gradually increase the total detector mass to 1000 kg while at the same time reducing the radioactive background in the ROI to $0.1\,\textrm{cts} / \textrm{(FWHM t yr)}$. LEGEND-1000 will be operated in a new experimental infrastructure which will allow the deployment and operation of several individual payloads.

\section{Signal Readout Electronics for \textsc{LEGEND}-200}
LEGEND-200 uses a resistive-feedback charge-sensitive amplifier (CSA) that is operated within the liquid argon (LAr) volume of the detector (see \cite{LEGEND} for additional details about the experimental setup). The CSA is separated into two stages in order to meet stringent radiopurity constraints.\\

\begin{wrapfigure}{r}{0.41\textwidth}
\centering
\includegraphics[width=0.4\textwidth]{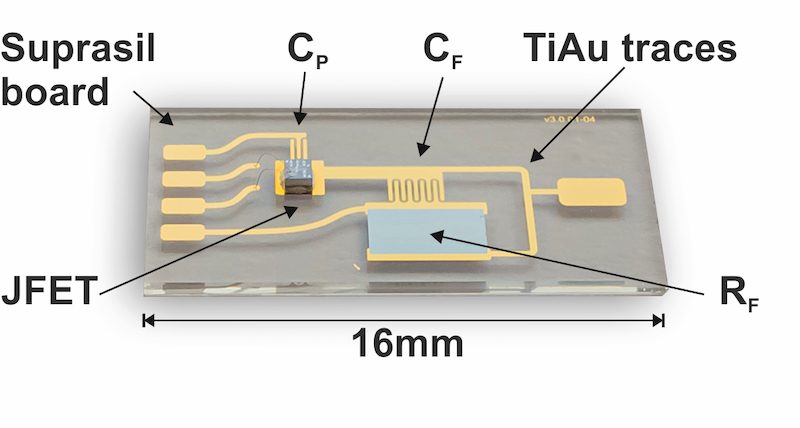}
\vspace{-5pt}
\caption{Image of a prototype LMFE device for LEGEND-200.}
\vspace{-5pt}
\label{fig:lmfe}
\end{wrapfigure}

\noindent The first stage -- the \textit{Low Mass Front End (LMFE)} -- is a custom low-mass, low-background, circuit which is located only a few mm away from each HPGe detector. It is based on the front end used in the \textsc{Majorana Demonstrator} experiment \cite{MJDFE}. A realised prototype device is shown in Figure \ref{fig:lmfe}. For each LMFE, only $\lesssim$ \SI{1}{\micro\becquerel} of radioactivity can be tolerated. Therefore, the LMFE is fabricated on an ultra-clean Suprasil substrate using titanium-gold (TiAu) traces, an in-die junction field effect transistor (JFET, Moxtek MX11) and an amorphous germanium thin-film feedback resistor ($\textrm{R}\textsubscript{F}\approx 1 - 2 \,\textrm{G}\Omega$ at 87 K). The a feedback capacitance ($\textrm{C}\textsubscript{F} \approx 400 \,\textrm{nF}$) and pulser capacitance ($\textrm{C}\textsubscript{P}\approx 100 \,\textrm{nF}$) for test-pulse injection are formed by the stray capacitance between the TiAu traces. Preliminary assay results indicate that the required radiopurity will be achieved.\\

\begin{figure}[h]
\centering
\includegraphics[width=0.85\textwidth]{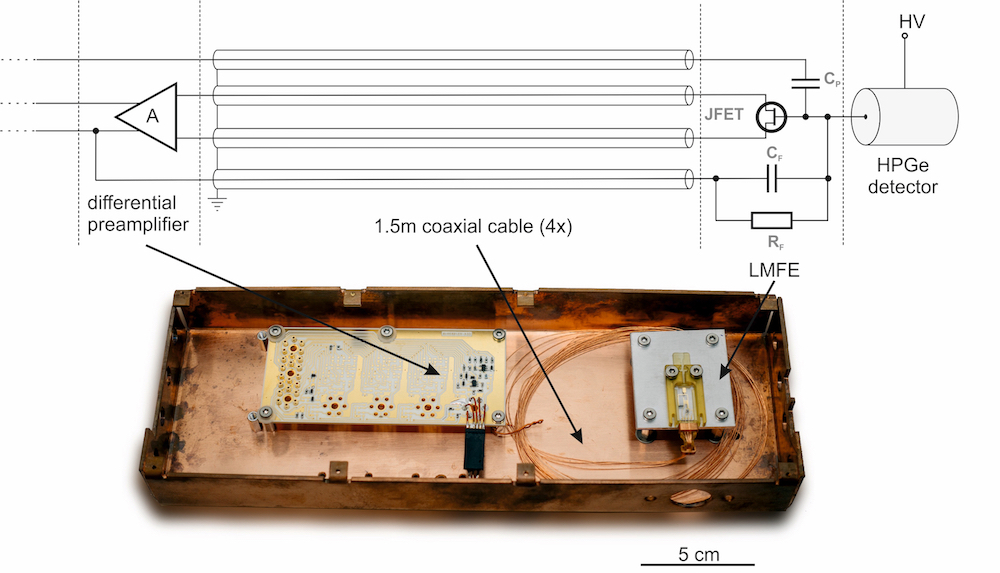}
\caption{\label{L200FEBench}Simplified circuit diagram of the LEGEND-200 signal readout electronics (top) and image of a realised prototype of the electronics used in bench-top measurements (bottom).}
\label{fig:csa}
\end{figure}

\noindent The second stage is a differential amplifier that is located at a distance of $\sim$ 30 - 150 cm away from the HPGe detector where a slightly higher radioactivity of \SI{50}{\micro\becquerel} per channel can be tolerated. The amplifier is based on the signal amplifier used in the GERDA experiment \cite{GERDA} and is fabricated using commercial off-the-shelf surface mount components on clean Kapton circuit boards. All components are assayed prior to fabrication in order to ensure the radiopurity goals are met.
The first and second stage are connected via 4 low-mass coaxial cables ({Axon picocoax}) which are custom made using specifically sourced and assayed copper wires and dielectric material. Figure \ref{fig:csa} shows a simplified circuit diagram of the CSA as well as an image of a prototype assembly used in benchtop measurements.\\

\noindent The CSA is designed with several key performance parameters in mind: a low electronic noise equivalent to $< 1 \,\textrm{keV}$ FWHM and an energy resolution of $\lesssim 2.5 \,\textrm{keV}$ in the ROI at $Q_{\beta\beta} = 2039 \,\textrm{keV}$; a fast rise time of $\lesssim 100 \,\textrm{ns}$ to allow pulse shape discrimination analysis of the signals; a high linearity of up to 10 MeV to detect high energy $\alpha$ decays which would provide information to model the background of the experiment; and the ability to drive a $\sim$ 10 m long transmission line between the differential amplifier and the data acquisition system.

\section{Summary}
LEGEND-200 is the first phase of the LEGEND experimental program and will operate 200 kg of HPGe detectors enriched in the isotope $^{76}\textrm{Ge}$ to search for $0\nu\beta\beta$ decay with a half-life sensitivity of $10^{27}\,\textrm{yr}$.
 A key aspect in reaching this unprecedented sensitivity is the reduction of radioactive contaminations in the vicinity of the detectors and the background reduction through pulse shape discrimination. Low-mass, low-background signal readout electronics are essential to achieve both aspects. LEGEND-200 uses a resistive feedback charge-sensitive amplifier, operated in LAr, that is based on previous successful deployments by the GERDA and \textsc{Majorana Demonstrator} experiments. On-site construction for LEGEND-200 will start in mid-2020 at the Gran Sasso underground laboratory in Italy.

\ack
This material is based upon work supported by the U.S. NSF, DOE-NP, NERSCC and through the LANL LDRD program, the Oak Ridge Leadership Computing Facility; the Russian RFBR, the Canadian NSERC and CFI; the German BMBF, DFG and MPG; the Italian INFN; the Polish NCN and Foundation for Polish Science; and the Swiss SNF; the Sanford Underground Research Facility, and the Laboratori Nazionali del Gran Sasso.\\ M. Willers gratefully acknowledges support by the Alexander von Humboldt Foundation.

\end{document}